\documentclass[PHME, 2026]{PHMSociety}

\usepackage{graphicx}
\usepackage{amsmath}
\usepackage{mlstyle}
\usepackage{amssymb}
\usepackage{subcaption}
\usepackage{siunitx}
\usepackage{tikz}
\usetikzlibrary{shapes.geometric,arrows.meta}
\usetikzlibrary{positioning}
\usepackage{makecell}
\usepackage{balance}
\usepackage{siunitx}

\begin{document}

\hyphenpenalty=1000
\exhyphenpenalty=1000

\title{Aging States Estimation and Monitoring Strategies of Li-Ion Batteries Using Incremental Capacity Analysis and Gaussian Process Regression}

\author{%
	Moritz Landwehr, Patrick Hoher and Johannes Reuter
}

\address{
	\affiliation{}{HTWG University of Applied Sciences, Konstanz, Germany}{ 
		{\email{moritz.landwehr@htwg-konstanz.de}}\\ 
        {\email{phoher@htwg-konstanz.de}}\\ 
		{\email{jreuter@htwg-konstanz.de}}
		} 
}

\maketitle
\pagestyle{fancy}
\thispagestyle{plain}

\phmLicenseFootnote{Moritz Landwehr}

\begin{abstract}
 Existing approaches for battery health forecasting often rely on extensive cycling histories and continuously monitored cells. In contrast, many real-world scenarios provide only sparse information, e.g. a single diagnostic cycle. In our study, we investigate \gls{SoH}- and \gls{RUL} estimation of previously unseen lithium-ion cells, relying on cycling data from \gls{BOL} to \gls{EOL} of multiple similar cells by using the publicly available Oxford battery aging dataset. The estimator applies \gls{ICA}-based feature extraction in combination with data-efficient regression methods. Particular emphasis is placed on a multi-model \gls{GPRn}, which also provides uncertainty quantification. Due to a rather cell invariant behaviour, the mapping of \gls{ICA} features to \gls{SoH} estimation is highly precise and points out a \gls{NMAE} of 1.3\%. The more cell variant mapping to \gls{RUL} estimation is challenging, reflecting in a \gls{NMAE} of 5.3\%.  Using the estimation results, a  \gls{RUL} monitoring strategy is derived. The objective is to safely operate a battery cell from \gls{BOL} to \gls{EOL} by only taking sparse diagnostic measurements. On average,  only four diagnostic measurements are required during a cell's lifetime of 3300 to 5000 cycles.
\end{abstract}


\section{Introduction}
\label{sec:Intro} 

\subsection{Motivation}
Lithium-ion batteries are a key technology for electric vehicles, stationary storage, and portable electronics and their aging directly impacts system reliability, safety, and cost. Capacity fade and resistance increase reduce usable energy and power, while uncontrolled degradation can lead to unplanned downtime and safety concerns. Accurate estimation of \gls{SoH} and \gls{RUL} is therefore essential for predictive maintenance, lifetime extension, and economic decision-making, especially in fleet and second-life applications. In many practical scenarios, however, only a few diagnostic cycles are available instead of continuous, long-term monitoring of individual cells. Methods that rely on dense cycling histories are thus difficult to apply, which creates a need for data-efficient approaches that extract information-rich features from sparse measurements and provide reliable \gls{SoH}- and \gls{RUL} estimates with quantified uncertainty.

\subsection{Battery Modelling and Estimation Methods}
In \cite{demirci_review_2024} and \cite{schmitt_state_2022} an overview about different methods of \gls{SoH} estimation is shown, whereas \gls{RUL} estimation methods are grouped and summarized in \cite{elmahallawy_comprehensive_2022}. Both, \gls{SoH}- and \gls{RUL} estimation, deal with long term dynamic behaviour of battery cells, which is in contrast to \gls{SoC} estimation. The methods for \gls{SoH}- and \gls{RUL} estimation can be grouped in model-based and experimental methods. Regarding model-based methods, battery aging can be modelled using a broad spectrum of approaches with different complexity, accuracy and computational intensity \cite{hamar_data-driven_2024}. Most widespread modelling approaches are electro-chemical models \cite{santhanagopalan_review_2006}, electric equivalent circuit models \cite{guo_review_2021}, electrochemical impedance models \cite{jossen_fundamentals_2006}, open circuit potential models, and data-driven models. For \gls{SoH} estimation there are also several experimental methods, dividing in direct measurement methods, like impedance measurement and coulomb counting, and indirect analysis methods, like \gls{ICA} and \gls{DVA} \shortcite{demirci_review_2024}. Electro-chemical models typically require extensive parametrization, detailed knowledge of internal cell chemistry, and significant computational effort, which limits their scalability and applicability across different cell types and operating conditions. For electric equivalent circuit models, the aging behaviour of the circuit elements has to be mathematically described and optimized using a large, high resolution dataset. Data-driven, specifically \gls{ML}-approaches, have therefore gained increasing attention \cite{dar_comprehensive_2025}.

\subsection{Machine Learning in Battery State Estimation}
\cite{al-hashimi_decade_2025} review, that different neural networks topologies, like \glspl{FFNN}, \glspl{LSTM}, \glspl{SVR}, \glspl{GPR}, and deep learning architectures have been successfully applied to \gls{SoH}- and \gls{RUL} estimation. Mainly voltage, current, temperature, and capacity over time from charging and discharging cycles are used as regressor inputs. They are comparably easy to measure. Intuitively, these features can be directly fed into one of the previously mentioned regressors. However it is a challenge to map the sparse information density of raw measurement data to meaningful features, describing the battery aging process. In the literature there are some examples given. In \cite{jha_itransformer_2024} a deep learning transformer network directly estimates the \gls{RUL} with time based features as inputs. Also \cite{nguyen_van_estimation_2023} processes time based input data with a \gls{LSTM} for \gls{SoH} estimation.  There are also regressors, like \glspl{CNN}, including inherently feature extraction \cite{zhou_battery_2025}. All this approaches are based on observing batteries from \gls{BOL} to a certain age as training data source, to afterwards predict the development of the aging until \gls{EOL}. In contrast, we deal with the task of a \gls{SoH}- and \gls{RUL} estimation of previously unseen cells using cycling data from \gls{BOL} to \gls{EOL} of multiple similar cells. In this case, due to cell-to-cell variance the direct processing of sparse information dense time based features is critical. Here \gls{ICA} has emerged as a powerful non-destructive diagnostic technique to extract high information dense aging-related features.

\subsection{Battery State Estimation Using \gls{ICA}} 
As \cite{she_battery_2023} describe, the \gls{ICA} method transforms voltage plateaus and inflection points on the voltage-capacity curves to identifiable peaks and valleys on the corresponding \gls{IC} curves. The characteristic shape of the \gls{IC} curves directly corresponds to degradation mechanisms, like \gls{LLI}, \gls{LAM} and \gls{ORI}. \gls{LLI} is generally the primary source, leading to direct capacity fade \cite{ansean_lithium-ion_2019}. So \gls{ICA} curves retain a certain degree of physical interpretability. Thanks to these compelling properties, during the last ten years \gls{SoH}- and \gls{RUL} estimation based on \gls{ICA} features have become popular. \cite{wang_state--health_2023} maps the extracted \gls{ICA} features simply by a low order polynomial fit to the \gls{SoH} estimation. They are selecting one or more features depending on charging or discharging cycle and the initial \gls{SoC}. A further advanced method is suggested in \cite{li_state_2020-1}, where the mapping from \gls{ICA} features to \gls{SoH} is done with a trained \gls{SVR}. In both papers, training data is also taken from similar cells, while the test cell is regarded as previously unseen. Also \gls{RUL} estimation is already done using \gls{ICA} features, \cite{xia_state_2023} apply a \gls{BiGRU} for \gls{RUL} estimation. \cite{pang_lithium-ion_2021} even use \gls{GPR} as regression method, both papers train single cell aging models with cycles from \gls{BOL} to a certain threshold and test for cycles from the threshold to \gls{EOL}. 

\subsection{Proposed Solution}
 In a typical application, a battery returns from an arbitrary unknown usage. After completing one (partial) charging cycle, \gls{SoH} and \gls{RUL} should be estimated under assessable uncertainties. Based on estimation and uncertainty, a decision on the further use can be taken. A proper method for uncertainty estimation in \gls{ML} regressors are \glspl{GP} \cite{rasmussen_gaussian_2008}. There are publications like \cite{li_state_2020}, also using \gls{ICA} and \gls{GPR} for \gls{SoH} estimation. However, they perform single cell estimation using the first 55\% of the cycles as training data to predict further development. In our study, we focus on a scenario where the regression maps one charging cycle of a previously unseen battery cell to the \gls{SoH}- and \gls{RUL} estimation, by knowing cycling data from \gls{BOL} to \gls{EOL} of multiple similar cells. The main contributions of this paper are
\begin{itemize}
    \item Derivation of a \gls{GPRn} regressor, combining single training cell \glspl{GP},
    \item Comparison of \gls{GPRn} performance to other standard \gls{ML} regressors, regarding  \gls{SoH}- and \gls{RUL} estimation,
    \item Derivation of a \gls{RUL} monitoring strategy based on \gls{GPRn} uncertainty estimation,
    \item Hyperparameter tuning and quantitative evaluation of the \gls{RUL} monitoring strategy.
\end{itemize}
The proposed \gls{RUL} monitoring strategy manages controlled aging until \gls{EOL} by only taking measurements on a few charging cycles during battery lifetime.

\subsection{Organization of the Paper}
The paper is organized as follows: Section 2 explains the \gls{ICA} process, coming from raw data to \gls{ICA} features. Further, it describes the theory of our \gls{GPRn} approach, serving as \gls{SoH}- and \gls{RUL} estimator and explains the RUL monitoring strategy. In section 3 the \gls{GPRn} based \gls{SoH}- and \gls{RUL} estimators are quantitively compared to other \gls{ML} estimators. Also  the result of the \gls{RUL} monitoring strategy is presented. Key conclusions and an outlook are drawn in section 4.

\section{Methods}
\label{sec:Methods}

\subsection{Oxford Battery Dataset}
\label{subsec:OBD}
In our study, the Oxford Battery Degradation Dataset 1 is used \cite{birkl_oxford_2017}. It comprises aging experiments conducted on eight small lithium-ion pouch cells with a nominal capacity of $740~\mathrm{mAh}$ and rated voltage of $3.6~\mathrm{V}$. Each cell employs a lithium cobalt oxide positive electrode and a graphite negative electrode. The tests were performed in a  temperature chamber maintained at $40~^\circ\mathrm{C}$, using a \gls{CC} and \gls{CV} charging protocol with upper and lower cut-off voltages of $4.2~\mathrm{V}$ and $2.6~\mathrm{V}$, respectively. The charging rate in \gls{CC} is 1C. Discharging is done with defined driving profiles. Voltage, charge, and cell temperature were recorded during diagnostic cycles conducted every 100 cycles at a sampling rate of $1\,\mathrm{Hz}$. The practical \gls{EOL} of batteries in mobile applications often is defined at 80\% \gls{SoH}. This is reached by the eight cells in between 3300 and 5300 cycles \cite{birkl_diagnosis_2016}.

\subsection{Incremental Capacity Analysis}
\label{subsec:ICA}

\begin{figure}[h!]
    \centering

    \begin{subfigure}{\linewidth}
        \centering
        \includegraphics[width=0.9\linewidth]{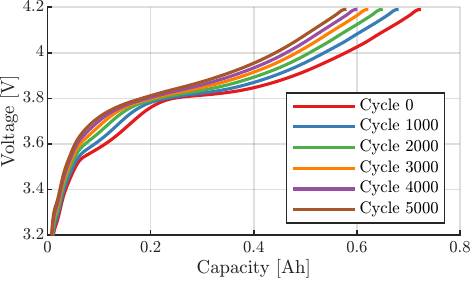}
        \caption{Standard charging curve}
        \label{fig:ic_a}
    \end{subfigure}

    \begin{subfigure}{\linewidth}
        \centering
        \includegraphics[width=0.9\linewidth]{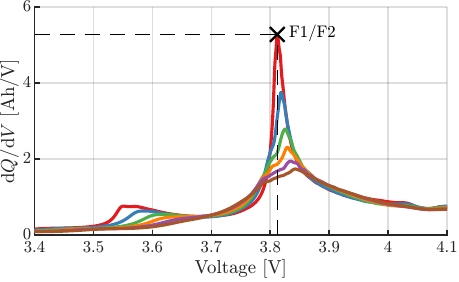}
        \caption{IC charging curve}
        \label{fig:ic_b}
    \end{subfigure}

    \begin{subfigure}{\linewidth}
        \centering
        \includegraphics[width=0.9\linewidth]{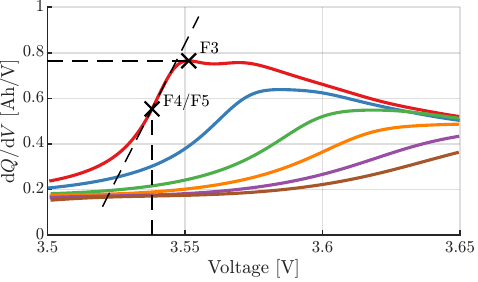}
        \caption{IC charging curve zoom}
        \label{fig:ic_c}
    \end{subfigure}

    \caption{Comparison of charging curves}
    \label{fig:comp_curves}
\end{figure}

With \gls{ICA}, battery aging mechanisms are analyzed using their peaks and valleys. The \gls{IC} is defined as shown in equation \ref{eq:IC_def}, where $Q$ represents the charge and $V$ the voltage across the battery terminals.
\begin{equation}
\label{eq:IC_def}
    \mathrm{IC} = \frac{\mathrm{d}Q}{\mathrm{d}V} = \frac{I \cdot \mathrm{d}t}{\mathrm{d}V} = I \cdot \frac{\mathrm{d}t}{\mathrm{d}V}
\end{equation}
So \gls{IC} in words means, how much charge is stored into the battery per voltage change. The \gls{ICA} method converts the voltage plateaus in the voltage curve into clearly identifiable peaks on the \gls{IC} curve \cite{ansean_lithium-ion_2019}. To illustrate the advantage of transforming the measurement data into an \gls{IC} curve, figure \ref{fig:comp_curves} compares the standard charging curve with the \gls{IC} curve of a single cell's life. 
 The comparison clearly shows better interpretability of cell aging in the \gls{IC} curve. Even if only partial charging segments are available (e.g. $3.5~\mathrm{V} < V_{\mathrm{meas}} < 3.9~\mathrm{V}$), relevant information for aging prognosis is still included. For the standard charging curve, the most visible indicator is the capacity at end of charge, which in fact is a direct representative of the \gls{SoH}, due to the definition \cite{yao_review_2021} 
 \begin{equation}
\label{eq:SoH_def}
   \mathrm{SoH} = \frac{Q_m}{Q_r} \cdot 100 \%,
\end{equation}
where $Q_m$ is the current maximum available capacity of the battery, measured under rated conditions, and $Q_r$ is the rated capacity. So the standard charging curve only provides highly interpretable information about cell aging when the battery is fully charged from $\mathrm{SoC} = 0$ to $\mathrm{SoC} = Q_m$.  

Normally, only the \gls{CC} segment of the charging cycle is considered, which implicitly defines $I$ in equation (\ref{eq:IC_def}). As shown in \cite{wang_state--health_2023} the charging rate C influences the distinctiveness of characteristic features of the \gls{IC} curve significantly. High currents heavily influences cell reactions and distort the \gls{IC} curve characteristic \cite{stroe_lithium-ion_2020}. When the current $I$ is fixed by the \gls{CC} charging protocol, only the time derivative of the voltage $\frac{\mathrm{d}V}{\mathrm{d}t}$ has to be calculated from raw measurement data. There is always an inevitable error and noise in battery operation and battery voltage measurement. Also, for a proper sampling frequency, the voltage change in between two time steps is $ \Delta V = V_{{k}} - V_{{k-1}} \approx 0$, making the direct numeric derivative $\frac{\mathrm{d}V}{\mathrm{d}t}$ of the measurement data extremely noisy. Filtering methods are needed to smooth \gls{IC} curves \cite{chen_filter_2024}. \cite{beatty_review_2024} reviews several filter methods, including moving average, gaussian filter, Kalman filter, butterworth low-pass filter and Savitzky–Golay filter. In our case a 4th degree butterworth low-pass with a cut frequency of $f_c = 0.01~\mathrm{Hz}$ is applied to filter the voltage measurement time series by using MATLAB's \textit{filtfilt()} zero-phase digital filtering. The resulting \gls{IC} curve is finally smoothed using a moving average filter. 

Features describing the aging of the battery cell have to be defined. For smaller datasets like the given one, feature definition is done manually by data observation and analysis. Looking at figure \ref{fig:comp_curves} b), the most obvious feature F1 is the maximum peak of the \gls{IC} curve $\mathrm{IC_{peak}}$, as it significantly decreases with further aging. Feature F2 is the corresponding voltage $V(\mathrm{IC_{peak}})$. Zooming into the interval $3.5~\mathrm{V} < V_{\mathrm{bat}} < 3.65~\mathrm{V}$ at figure \ref{fig:comp_curves} c), three more features are derived: F3 is the maximum occurring \gls{IC}, named $\mathrm{IC_{max}}$ for this given interval. F4 is the maximum positive slope of the \gls{IC} curve in this interval, ${\frac{\mathrm{d}\mathrm{IC}}{\mathrm{d}V}}_{\mathrm{max}}$. F5 is is the corresponding voltage $V({\frac{\mathrm{d}\mathrm{IC}}{dV}}_{\mathrm{max}})$, where F4 occurs.

\begin{figure}[ht]
    \centering
    \begin{subfigure}{\linewidth}
        \centering
        \includegraphics[width=0.9\linewidth]{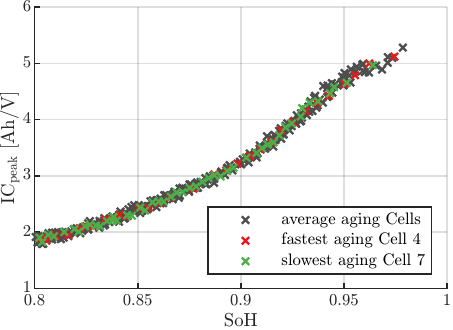}
        \caption{\gls{SoH} cell variance}
        \label{fig:comp_a}
    \end{subfigure}

    \begin{subfigure}{\linewidth}
        \centering
        \includegraphics[width=0.9\linewidth]{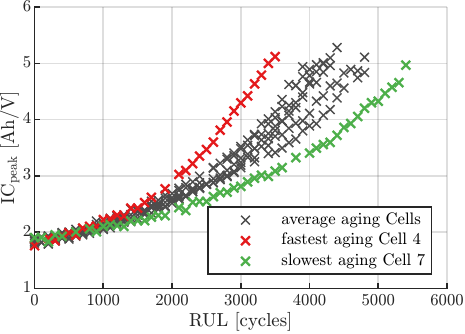}
        \caption{\gls{RUL} cell variance}
        \label{fig:comp_b}
    \end{subfigure}

    \caption{Cell variance of feature F1 regarding \gls{SoH} and \gls{RUL}}
    \label{fig:cell_vairance_soh_rul}
\end{figure}

The chosen features are evaluated using the Spearman rank correlation coefficient with the single feature as input quantity and the true \gls{SoH} and \gls{RUL} as output quantity. \gls{RUL} is defined as the number of remaining charge--discharge cycles, until the battery's \gls{SoH} reaches the \gls{EOL} threshold of
$80\,\%$, i.e.,
\begin{equation}
\label{eq:RUL_def}
\mathrm{RUL} = N_{\mathrm{EOL}} - N_{\mathrm{current}}, 
\quad \text{with} \quad
\mathrm{SoH}_{\mathrm{EOL}} = 0.8 .
\end{equation}
The Spearman rank correlation coefficient is a non-parametric measure of a monotonic association between two variables \cite{Spearman}. It  can be expressed as
\begin{equation}
\label{eq:Spearman_def}
\rho_s
= 1 - \frac{6 \sum_{i=1}^{n} d_i^2}{n(n^2-1)},
\end{equation}
where $d_i = R[X_i] - R[Y_i]$ is the difference in ranks (sorted positions) of the $i$-th data pair and $n$ is the number of observations. Unlike the Pearson coefficient, Spearman’s $\rho_s$ does not assume linearity or normality and is therefore robust for assessing monotonic but non-linear relationships.
\begin{table}[ht]
\centering
\caption{Spearman rank correlation coefficients of IC features}
\label{tab:correlation}
\renewcommand{\arraystretch}{1.15}
\begin{tabular}{c|c|c|c|c|c}
\hline
 & \textbf{F1} & \textbf{F2} & \textbf{F3} & \textbf{F4} & \textbf{F5} \\
\hline   
\textbf{SoH} & 0.9974 & -0.9817 & 0.9873 & 0.9830 & -0.9908 \\
\textbf{RUL} & 0.9702 & -0.9488 & 0.9814 & 0.9811 & -0.9715 \\
\hline
\end{tabular}
\end{table}
Shown in table \ref{tab:correlation}, the correlation of a single feature to \gls{SoH} is always greater than the correlation to \gls{RUL}. This is due to higher cell-to-cell variances in mapping the features to \gls{RUL} than to \gls{SoH}. Figure \ref{fig:cell_vairance_soh_rul} shows the mapping of the dominant feature $\mathrm{IC_{peak}}$ to the true \gls{SoH} and \gls{RUL} respectively.
\subsection{Gaussian Process Regression}
\label{subsec:GPR}
\gls{GPR} is used in this work to probabilistically model the nonlinear relation between \gls{ICA}-based features and the aging quantities \gls{SoH} and \gls{RUL}. A Gaussian Process defines a distribution over functions
\begin{equation}
    f(\mathbf{x}) \sim \mathcal{GP}\big(m(\mathbf{x}), k(\mathbf{x},\mathbf{x}')\big),
\end{equation}
which is fully specified by the mean function $m(\mathbf{x})$ and the covariance function $k(\mathbf{x},\mathbf{x}')$ \cite{rasmussen_gaussian_2008}. 

For the \gls{ICA} feature vectors $\mathbf{x} \in \mathbb{R}^d$, a \gls{RBF}, squared exponential kernel for d-dimensional feature is employed
\begin{equation}
    k(\mathbf{x},\mathbf{x}') 
    = \sigma_f^2 \exp\!\left(
        -\frac{1}{2}
        \sum_{j=1}^{d}
        \frac{\big(\mathbf{x}_j - \mathbf{x}'_j\big)^2}{\ell_j^2}
    \right).
\end{equation}
$\sigma_f^2$ denotes the signal variance and $\ell_j$ are the characteristic length scales of the individual \gls{ICA} features \cite{li_state_2020}. By setting individual length scales for every feature, \gls{ARD} defines which features dominate the regression performance.

The observed target quantities of all $n$ training samples, i.e. \gls{SoH} or \gls{RUL}, are denoted as $y_i$. They are modelled as noisy function evaluations from the hypothetical underlying latent function $f(\cdot)$
\begin{equation}
    y_i = f(\mathbf{x}_i) + \varepsilon_i, 
    \quad
    \varepsilon_i \sim \mathcal{N}(0,\sigma_n^2),  
    \quad 
    i = 1,...,n.
\end{equation}
The prior distribution can be denoted as
\begin{equation}
    \mathbf{y} \sim \mathcal{N}\big(
        m(\mathbf{x}),\,
        \mathbf{K}(\mathbf{x},\mathbf{x}) + \sigma_n^2 \mathbf{I}
    \big),
\end{equation}
with the n-dimensional symmetric positive definite kernel matrix $\mathbf{K}$ elements $\mathbf{K}_{ij} = k(\mathbf{x}_i,\mathbf{x}_j)$. For a new \gls{ICA} feature vector $\mathbf{x}_*$, the joint prior distribution of $\mathbf{y}$ and $f_* = f(\mathbf{x}_*)$ is given by
\begin{equation}
    \begin{bmatrix}
        \mathbf{y} \\
        f_*
    \end{bmatrix}
    \sim
    \mathcal{N}\!\left(
        m(\mathbf{x}),
        \begin{bmatrix}
            \mathbf{K}(\mathbf{x},\mathbf{x}) + \sigma_n^2 \mathbf{I} & \mathbf{k}_* \\
            \mathbf{k}_*^\top & k_{**}
        \end{bmatrix}
    \right),
\end{equation}
with $\mathbf{k}_* = [\,k(\mathbf{x}_1,\mathbf{x}_*),\dots,k(\mathbf{x}_n,\mathbf{x}_*)\,]^\top$ and $k_{**} = k(\mathbf{x}_*,\mathbf{x}_*)$. From that, the posterior distribution of the test output data $\mathbf{f}^*$, knowing training data $\mathbf{x}$, $\mathbf{y}$ and test input features $\mathbf{x}^*$, is 
\begin{equation}
p(\mathbf{f}^* \mid \mathbf{x}, \mathbf{y}, \mathbf{x}^*)
= \mathcal{N}\big(  \bar{\mathbf{f}}^*, cov(\mathbf{f}^*) \big),
\label{eq:gpr_pred_dist}
\end{equation}
where the prediction mean $\bar{\mathbf{f}}^*$ and the prediction covariance $cov(\mathbf{f}^*)$ are given by
\begin{equation}
\bar{\mathbf{f}}^*
= \mathbf{K}(\mathbf{x},\mathbf{x}^*)^{\mathrm{T}}
\left[
\mathbf{K}(\mathbf{x},\mathbf{x}) + \sigma_n^2 \mathbf{I}
\right]^{-1}
\mathbf{y}
\label{eq:gpr_pred_mean}
\end{equation}
and
\begin{align}
cov(\mathbf{f}^*)
&= \mathbf{K}(\mathbf{x}^*,\mathbf{x}^*) \nonumber \\
&- \mathbf{K}(\mathbf{x},\mathbf{x}^*)^{\mathrm{T}}
\left[
\mathbf{K}(\mathbf{x},\mathbf{x}) + \sigma_n^2 \mathbf{I}
\right]^{-1}
\mathbf{K}(\mathbf{x},\mathbf{x}^*).
\label{eq:gpr_pred_var}
\end{align}
The hyperparameter set
\(
\boldsymbol{\theta}
= 
\big\{
\sigma_f^2,\,
\ell_1,\dots,\ell_d,\,
\sigma_n^2
\big\}
\)
is obtained by maximizing the \gls{LML}
\begin{align}
    \log p(\mathbf{y}\mid\mathbf{x},\boldsymbol{\theta})
    &=
    -\frac{1}{2}\,
    \mathbf{y}^\top
    \big(\mathbf{K}(\mathbf{x},\mathbf{x}) + \sigma_n^2 \mathbf{I}\big)^{-1}
    \mathbf{y} \nonumber \\
    &-\frac{1}{2}\,
    \log\!\big|\mathbf{K}(\mathbf{x},\mathbf{x}) + \sigma_n^2 \mathbf{I}\big|
    -\frac{n}{2}\log(2\pi),
\end{align}
using gradient-based optimization with Pythons \textit{Adam()} optimization environment. 

\subsection{\gls{GPR} Ensemble Approach \gls{GPRn}}
A single \gls{GPR} model trained on pooled data from all cells cannot capture the pronounced cell-to-cell variability observed in the Oxford dataset, in particular for the mapping from \gls{ICA} features to \gls{RUL}. To address this, an ensemble of cell-specific \glspl{GPR}, denoted as \gls{GPRn}, has been developed in this work. For each of the $T$ training cells, a separate \gls{GP} expert is fitted, using only the ICA feature vectors and target labels (SoH or RUL) of that particular cell. Assume, that for a new feature vector $\mathbf{x}^*$, each of the $T$ \gls{GP} experts provides a univariate Gaussian predictive distribution
\begin{equation}
    p_i(y \mid \mathbf{x}^*) = \mathcal{N}\big(m_i, s_i^2\big), 
    \qquad i = 1,\dots,T,
\end{equation}
with mean $m_i$ and variance $s_i^2$. The \gls{GPRn} ensemble combines these experts as a finite mixture of Gaussians with non- negative weights $\pi_i$ that sum to one, $\sum_{i=1}^T \pi_i = 1$,
\begin{equation}
\label{eq:pi}
    p(y \mid \mathbf{x}^*) 
    = \sum_{i=1}^T \pi_i \,
      \mathcal{N}\big(m_i, s_i^2\big).
\end{equation}
The predicted mean of the mixture is the expectation of $y$ under $p(y \mid \mathbf{x}^*)$,
\begin{equation}
    \mu 
    = \mathbb{E}[y]
    = \sum_{i=1}^T \pi_i \, \mathbb{E}_i[y]
    = \sum_{i=1}^T \pi_i m_i.
\end{equation}
The predicted variance is obtained from variance definition equation
\begin{equation}
    \operatorname{Var}(y) 
    = \mathbb{E}[y^2] - \big(\mathbb{E}[y]\big)^2.
\end{equation}
After some simple manipulations the variance results in
\begin{align}
\label{eq:Variance_epi_alea}
    \operatorname{Var}(y)
    &= \mathbb{E}[y^2] - \mu^2 \nonumber\\
    &= \sum_{i=1}^T \pi_i \big(s_i^2 + m_i^2\big)
       - \left(\sum_{i=1}^T \pi_i m_i\right)^2 \nonumber\\
    &= \underbrace{\sum_{i=1}^T \pi_i s_i^2}_{\text{epistemic uncertainty}} +
       \underbrace{\left(
         \sum_{i=1}^T \pi_i m_i^2
         - \left(\sum_{i=1}^T \pi_i m_i\right)^2
       \right)}_{\text{aleatoric uncertainty}}.
\end{align}
The first term corresponds to the average predictive variance of the individual experts and captures the epistemic uncertainty of the data. It is caused by a lack of knowledge, because information about the true \gls{SoH} and \gls{RUL} are only available at finite samples. The second term is the variance of the single \gls{GP} expert means and measures how strongly the experts disagree. This is interpreted as aleatoric uncertainty arising from inherent random cell-to-cell variance.

To prevent the single \gls{GP}s from overfitting, the number of training epochs $n$ of the \gls{GPR} is restricted. If the number of epochs is too large, single \gls{GP}s overfit, resulting in a higher \gls{MAE} when evaluating the \gls{GPRn} approach with training data. Therefore $n$ is set, where the \gls{MAE} over $n$ is at its minimum.

\subsection{RUL Monitoring Strategy}
\label{subsec:CMS}
In real-world applications, including second-life applications, continuous cell monitoring is not always guaranteed, respectively expensive. Our proposed \gls{RUL} monitoring strategy enables informed operational decisions based on sparse measurements, typically three to five diagnostic cycles over the entire battery lifetime. The procedure leverages the \gls{GPRn} based variance including \gls{RUL} estimation and the \gls{SVR} based \gls{SoH} estimation to balance safe operation with maximal utilization.
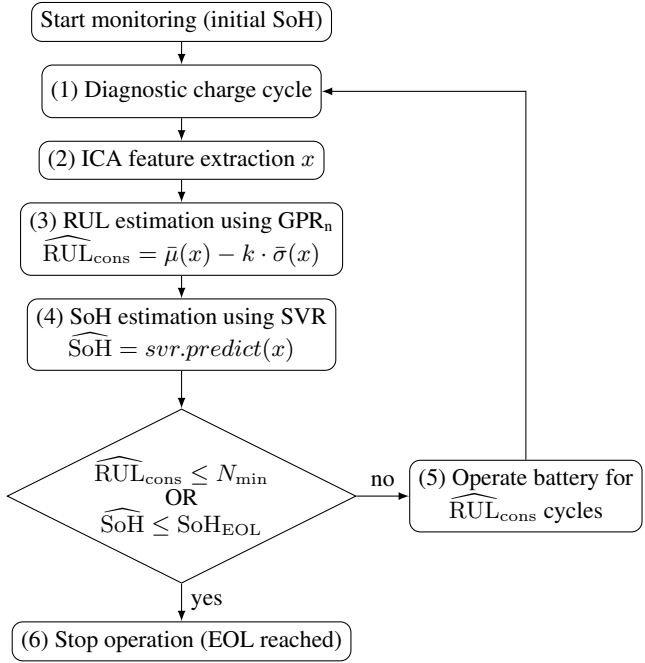
\begin{figure}[ht]
    \centering
    \begin{tikzpicture}[
        node distance=3mm and 3mm,
        every node/.style={font=\small},
        block/.style={rectangle, draw, rounded corners, align=center,
                      minimum width=10mm, minimum height=5mm},
        decision/.style={diamond, draw, aspect=2, align=center, inner sep=1pt},
        io/.style={rectangle, draw, rounded corners, align=center,
                  minimum width=32mm, minimum height=7mm},
        line/.style={->,>=latex}
    ]
    
    \node[block] (start) {Start monitoring (initial SoH)};
    \node[io,    below=of start] (meas) {(1) Diagnostic charge cycle};
    \node[block, below=of meas]  (ica)  {(2) ICA feature extraction $x$};
    \node[block, below=of ica]   (gprn) {(3) \gls{RUL} estimation using \gls{GPRn} \\  $\widehat{\mathrm{RUL}}_{\mathrm{cons}} = \bar{\mu}(x) - k\cdot\bar{\sigma}(x)$};
    \node[block, below=of gprn]  (rul)  {(4) \gls{SoH} estimation using \gls{SVR}\\
      $\widehat{\mathrm{SoH}} = svr.predict(x)$};
    \node[decision, below=of rul, yshift=-2mm] (check)
      {{$\widehat{\mathrm{RUL}}_{\mathrm{cons}} \le {{N}}_{\mathrm{min}}$}\\ OR\\ $ \widehat{\mathrm{SoH}} \le \mathrm{SoH}_{\mathrm{EOL}}$};
    \node[block, right=7mm of check] (operate)
      {(5) Operate battery for\\$\widehat{\mathrm{RUL}}_{\mathrm{cons}}$ cycles};
    \node[block, below=of check, yshift=-2mm] (stop)
      {(6) Stop operation (EOL reached)};
    
    \draw[line] (start) -- (meas);
    \draw[line] (meas)  -- (ica);
    \draw[line] (ica)   -- (gprn);
    \draw[line] (gprn)  -- (rul);
    \draw[line] (rul)   -- (check);
    
    \draw[line] (check) -- node[right]{yes} (stop);
    \draw[line] (check.east) -- node[above]{no} (operate.west);
    \draw[line] (operate.north) |- (meas.east);
    
    \end{tikzpicture}
    \caption{Program flow chart of the iterative \gls{RUL} monitoring strategy using \gls{ICA} features and the \gls{GPRn} ensemble.}
    \label{fig:PAP}
\end{figure}
The monitoring workflow is illustrated in figure \ref{fig:PAP} and operates iteratively: (1)~perform a single diagnostic charging cycle at arbitrary points in the battery's lifetime, (2)~extract \gls{ICA} features and feed them into the trained \gls{GPRn} ensemble, (3)~compute a conservative \gls{RUL} estimate by subtracting a margin proportional to the uncertainty, (4)~estimate the current \gls{SoH} using a \gls{SVR} model, (5)~operate exactly $\widehat{\mathrm{RUL}}_{\mathrm{cons}}$ cycles before the next measurement, (6)~repeat until the battery approaches \gls{EOL}. \gls{EOL} is defined by $\widehat{\mathrm{RUL}}_{\mathrm{cons}}\le {N}_{\mathrm{min}}$, or if the estimated \gls{SoH} is already smaller than $ \mathrm{SoH}_{\mathrm{EOL}}$. With $\mathrm{N}_{\mathrm{min}}$, it is possible to define application-specific values for the \gls{RUL}, at which a further diagnostic cycle is worthwhile. In this case, ${N}_{\mathrm{min}}=40$ was selected.  This adaptive scheme allows operators to tolerate uncertainty quantification and trigger additional diagnostics only, when uncertainty becomes prohibitively large.

The following \glspl{KPI} assess the effectiveness of the monitoring strategy across all cells:
\begin{itemize}
    \item \textbf{Utilization Grade} \\
    \(
    U = \frac{N_{\mathrm{cycles,strategy}}}{N_{\mathrm{cycles,possible}}},
    \)
    measures the percentage of available cycles actually used before triggering EOL. Higher values indicate effective exploitation of remaining capacity.
    
    \item \textbf{Number of Monitoring Steps}\\ 
    $M$, counts how many diagnostic cycles were required to reach EOL. Lower $M$ reduces measurement burden.
    
    \item \textbf{Overcycling Rate}\\
    $P_{\mathrm{over}}$, fraction of cells that fall below the SoH~$= 0.80$ threshold. Ideally $P_{\mathrm{over}} = 0$ to avoid cell damage.
    
    \item \textbf{End-of-Life Cycle Deviation}\\ 
    $\Delta {N}_{\mathrm{EOL}} = {N}_{\mathrm{true,EOL}} - {N}_{\mathrm{strategy,EOL}}$,
    absolute error in cycle count at EOL. Small deviations demonstrate predictive accuracy.
    
    \item \textbf{End-of-Life SoH Deviation}\\
    $\Delta \mathrm{SoH}_{\mathrm{EOL}} = \mathrm{SoH}_{\mathrm{strategy,EOL}} - \mathrm{SoH}_{\mathrm{true,EOL}}$,
    residual SoH error at strategy termination. Ideally positive and close to zero.
\end{itemize}
The interplay between these KPIs can be tuned by setting hyperparameter $k$ and reveals the trade-off between aggressive utilization and safe operation. A highly conservative strategy (large $k$) yields high safety but low utilization, conversely, an aggressive strategy risks overcycling. Therefore, optimal tuning is essential for practical deployment.

\section{Results and Discussion}
\label{sec:Results}

 \subsection{SoH and RUL Regression Setup}
\label{subsec:SoH_RUL_Results}
Different regression approaches are benchmarked for \gls{SoH} and \gls{RUL} prediction using the same \gls{ICA} feature set and a common cross-cell evaluation protocol. For each experiment, six cells are used for training and two cells for testing, and all $\binom{8}{2}$ train-test combinations are evaluated. Performance is reported as \gls{MAE} averaged over all train-test combinations and absolute maximum error occurring.

The following commonly used regression models are benchmarked against the shown \gls{GPR} approach from literature and our novel \gls{GPRn} approach:
\begin{itemize}
    \item \textbf{Poly1D}
    \begin{itemize}
        \item One-dimensional polynomial regression using only the dominant feature IC\textsubscript{peak} as input. Hyperparameter: Polynomial degree = 3.
    \end{itemize}

    \item \textbf{PolyMulti}
    \begin{itemize}
        \item Multivariate polynomial regression over the full \gls{ICA} feature vector. Hyperparameter: Polynomial degree = 3, feature interaction terms degree = 2.
    \end{itemize}

    \item \textbf{FFNN}
    \begin{itemize}
        \item Hyperparameter: Two hidden layers with 64 neurons each; ReLU activations, trained with \textit{Adam()}, 1000 epochs, learning rate $10^{-3}$.
    \end{itemize}


    \item \textbf{SVR}
    \begin{itemize}
        \item \gls{SVR} with \gls{RBF} kernel and manually tuned hyperparameter: Regularisation parameter $C$, tube width $\varepsilon$ and feature-wise kernel length scales chosen by trial-and-error based on T-fold cross-validation loss.
    \end{itemize}
    
    \item $\text{\textbf{GPR}}_{\text{\textbf{LOCO}}}$
    \begin{itemize}
        \item Similar model and hyperparameter as \gls{GPR}, hyperparameter optimization by minimizing \gls{LOCO} validation loss.
    \end{itemize}


\end{itemize}

\subsection{SoH Regression Results}
For \gls{SoH} estimation, the dataset extends \gls{ICA} features from $\mathrm{SoH} = 100\%$ to $\mathrm{SoH} = 80\%$ for all eight cells. The \gls{SoH} results in table \ref{tab:SOH_results_overview} show that all investigated regression models achieve comparatively small average errors, but differ substantially in robustness and worst-case behaviour.
\begin{table}[h]
    \centering
    \caption{Performance metrics of \gls{SoH} estimation}
    \label{tab:SOH_results_overview}
    \renewcommand{\arraystretch}{1.15}
    \begin{tabular}{l|c|c|c}
        \hline
         & \makecell{\textbf{MAE train}\\ in \%}  & \makecell{\textbf{MAE test}\\ in \%} & \makecell{\textbf{Max Error test}\\ in \%} \\
        \hline
        \textbf{Poly1D} & 0.723 & 0.764 & 3.02 \\
        \textbf{PolyMulti} & 0.159 & 0.341 & 4.03 \\
        \hline
        \textbf{FFNN} & 0.220 & 0.305 & 1.66 \\
        \textbf{SVR} & 0.184 & 0.261 & 1.01 \\
        \textbf{GPR} & 0.214 & 0.271 & 1.20 \\
        \hline
        $\text{\textbf{GPR}}_{\text{\textbf{LOCO}}}$ & 0.202 & 0.317 & 5.64 \\
        \textbf{\gls{GPRn}} & 0.412 & 0.463 & 1.93 \\        
        \hline
    \end{tabular}    
\end{table}
The Polynomial fitting approaches provide a reasonable reference \gls{MAE}, but still exhibit relatively large maximum deviations, making them mainly a simple baseline. The commonly used \gls{ML} methods FFNN, SVR and GPR outperform polynomial fitting in lower \gls{MAE} and lower maximum error. SVR is evaluated as best performing method. Obviously, \gls{GPRn} performs worse than \gls{GPR}.  This can be explained by only minor cell-to-cell variance when \gls{ICA} features are mapped to SoH, as shown in figure \ref{fig:cell_vairance_soh_rul}. The single pooled GPR therefore uses approximately six times more data points to describe the SoH–feature relationship, reducing epistemic uncertainties.
\begin{figure}[ht]
     \centering
     \includegraphics[width=0.9\linewidth]{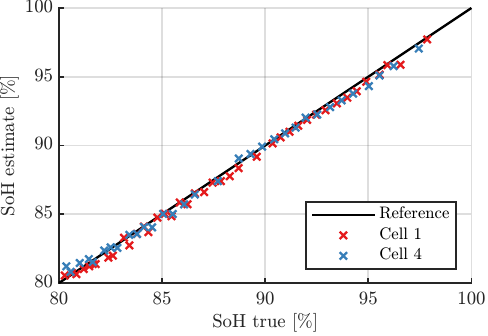}
     \caption{SoH regression example using SVR}
    \label{fig:SoH_reg_ex}
\end{figure} 
Figure \ref{fig:SoH_reg_ex} illustrates the \gls{SoH} regression using the best performing SVR approach for two test cells (Cell 1 and Cell 4) of the Oxford dataset. The regression results are highly satisfactory for both cells. 

\subsection{RUL Regression Results}
For \gls{RUL} estimation, the dataset extends \gls{ICA} features from $N_{\mathrm{cycles}} = 0$ to $N_{\mathrm{cycles}} = N_{\mathrm{EOL}} + 400$ for all eight cells. 
\begin{table}[ht]
    \centering
    \caption{Performance metrics of \gls{RUL} estimation}
    \label{tab:RUL_results_overview}
    \renewcommand{\arraystretch}{1.15}
    \begin{tabular}{l|c|c|c}
        \hline
         & \makecell{\textbf{MAE train}\\ in cycles}  & \makecell{\textbf{MAE test}\\ in cycles} & \makecell{\textbf{Max Error test}\\ in cycles} \\
        \hline
        \textbf{Poly1D} & 335 & 375 & 1468 \\
        \textbf{PolyMulti} & 111 & 299 & 2770 \\
        \hline
        \textbf{FFNN} & 226 & 247 & 1729 \\
        \textbf{SVR} & 236 & 271 & 1390 \\
        \textbf{GPR} & 36 & 323 & 2748 \\
        \hline
        $\text{\textbf{GPR}}_{\text{\textbf{LOCO}}}$ & 133 & 249 & 1734 \\
        \textbf{\gls{GPRn}} & 201 & 232 & 1277 \\
        \hline
    \end{tabular}    
\end{table}
In contrast to the SoH estimation, the RUL results in table \ref{tab:RUL_results_overview} highlight that direct remaining-life prediction is strongly affected by cell-to-cell variability. Cell-to-cell variability is averaged best by \gls{GPRn}, also \gls{FFNN} and \gls{SVR} perform satisfying. Here, the pooled GPR completely overfits, which can be seen at the extreme low MAE train compared to the MAE test. Already $\text{{GPR}}_{\text{{LOCO}}}$ reduces the overfitting by the cross validation mechanism, but still there are single outliers causing a high maximum error in the test dataset. The \gls{GPRn} ensemble, which explicitly models the cell-to-cell variability, gets rid of overfitting. This leads to noticeably lower mean and maximum RUL errors across all train-test splits. Additionally, the ensemble structure of \gls{GPRn} provides the uncertainty components, which are later exploited by the RUL monitoring strategy. 
\begin{figure}[h]
    \centering
    \begin{subfigure}{\linewidth}
        \centering
        \includegraphics[width=0.9\linewidth]{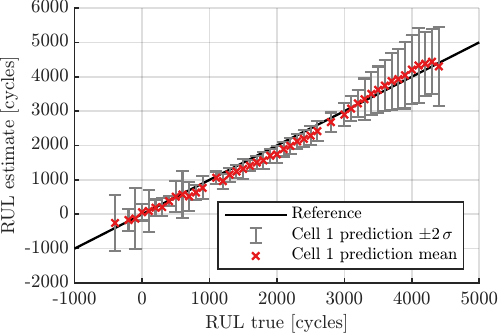}
        \caption{Cell 1}
        \label{fig:RUL_reg_1}
    \end{subfigure}

    \begin{subfigure}{\linewidth}
        \centering
        \includegraphics[width=0.9\linewidth]{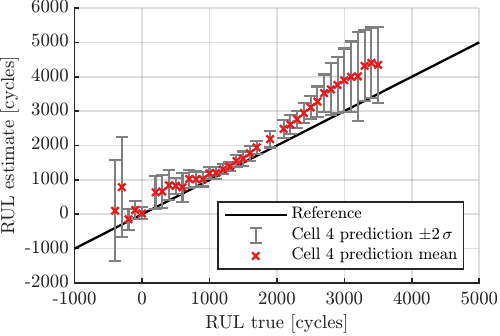}
        \caption{Cell 4}
        \label{fig:RUL_reg_4}
    \end{subfigure}

    \caption{RUL regression example using \gls{GPRn}}
    \label{fig:RUL_reg_ex}
\end{figure}
Figure \ref{fig:RUL_reg_ex} shows the \gls{RUL} regression using the best performing \gls{GPRn} approach for the same test cells as in the \gls{SoH} regression evaluation shown in figure \ref{fig:SoH_reg_ex}. The regression results are highly cell depended. For Cell 1, the \gls{RUL} estimation is performing much better than for Cell 4. Looking at figure \ref{fig:cell_vairance_soh_rul} b), one becomes aware, that Cell 1 has approximately average aging behaviour, whereas Cell 4 is aging fastest. Due to the averaging effect of \gls{GPRn}, the \gls{RUL} estimation performance improves, as more the test cell behaves like the average cell in the dataset.  


 \subsection{Comparsion of SoH and RUL Regression} 
To compare \gls{SoH} estimation with \gls{RUL} estimation a new measure has to be introduced, the \glsentrylong{NMAE} (\glsentryshort{NMAE}). It is defined by the mean absolute error divided by the range of the true values \cite{goldberg_eigentaste_2001}. The resulting metric becomes unitless, like for \gls{SoH} estimation
\begin{equation}
\label{eq_NMAE}
    \mathrm{NMAE}_{\mathrm{SoH}} =
    \frac{1}{N}
    \sum_{i=1}^{N}
    \frac{\left| \widehat{\mathrm{SoH}}_{i} - \mathrm{SoH}_{i} \right|}
         {\mathrm{SoH}_{\max} - \mathrm{SoH}_{\min}}.
\end{equation}
Using \gls{NMAE} allows a direct numerical comparison of \gls{SoH} and \gls{RUL} regression performance, even though \gls{SoH} is expressed in \% and \gls{RUL} in absolute cycle counts.
The cell-to-cell variances makes mapping of \gls{ICA} features to \gls{RUL} more challenging, resulting in a \gls{NMAE} of 5.33\%, which is around four times higher than for \gls{SoH} (1.31\%).

 \subsection{RUL Monitoring Strategy}
 \label{subsec:RUL_strategy_Results}
\begin{figure}[h!]
     \centering
     \includegraphics[width=0.9\linewidth]{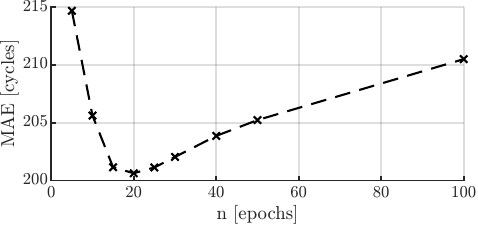}
     \caption{Tuning number of epochs based on MAE trraining loss}
    \label{fig:Hyperparameter_n}
\end{figure}

 Based on the results of \gls{RUL} estimation with \gls{GPRn} regression, the derived monitoring strategy is evaluated. The monitoring strategy basically inherits two tunable parameters. First, the number of epochs $n$ of the single \gls{GP}s training. It is defined finding a minimum in the training data \gls{MAE} over all train-test combinations. Figure \ref{fig:Hyperparameter_n} shows, that the minimum is located at $n \approx 20$.    
 \begin{figure}[h!]
     \centering
     \includegraphics[width=0.9\linewidth]{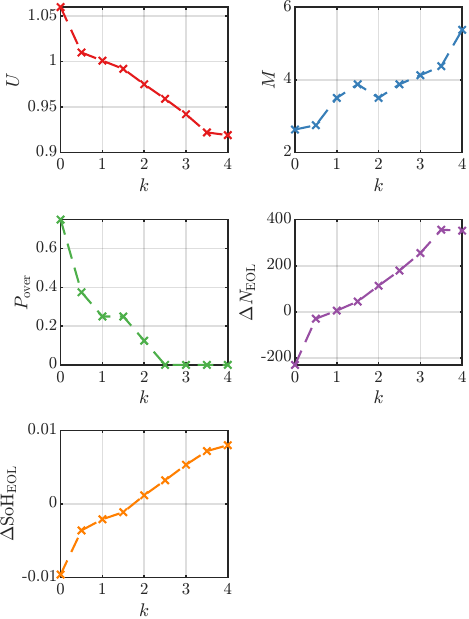}
     \caption{Tuning hyperparameter $k$ based on \glspl{KPI}}
    \label{fig:Hyperparameter_k}
\end{figure}

The second tunable parameter is $k$, trading-off between aggressive
utilization and safe operation. All in section \ref{subsec:CMS} defined \glspl{KPI} are effected from the selection of $k$, illustrated in figure \ref{fig:Hyperparameter_k}. In these five plots, mean results of the \glspl{KPI} over all train-test combinations are presented. Using low $k$, the utilization grade is set aggressive, resulting in $U > 1$, meaning in more than half of the examples the battery gets over-cycled ($P_{\mathrm{over}} > 0.5$). Therefore $\Delta {N}_{\mathrm{EOL}}$ also points out negative values. However, the number of monitoring steps $M \approx 3$ is rather small. With increasing $k$, the overcycling rate $P_{\mathrm{over}}$ consequently reduces due to focus on safer operation. $M$ therefore rises. A smooth trade-off is found for $k = 2$. Here $\Delta {N}_{\mathrm{EOL}}$ has a small, positive value, the overcycling rate $P_{\mathrm{over}} = \frac{1}{8}$ is already sufficiently small and $M=3.5$ still compromises a low diagnostic effort. Looking at figure \ref{fig:Hyperparameter_k}, the overcycling in terms of \gls{SoH} is limited for all chosen $k$ to $\Delta \mathrm{SoH}_{\mathrm{EOL}} \le \pm 1 \%$. Even the maximum overcycling of a single train-test combination for $k=0$ is $\mathrm{SoH}_{\mathrm{EOL}} = - 3.6 \%$, meaning no battery is aged to values $\mathrm{SoH} \le 76.4 \%$. Therefore, all selections of $k$ can be described as safe operation modes, which leaves the choice up to the user. For further presented examples $k=2$ is used.
 \begin{figure}[h!]
     \centering
     \includegraphics[width=0.9\linewidth]{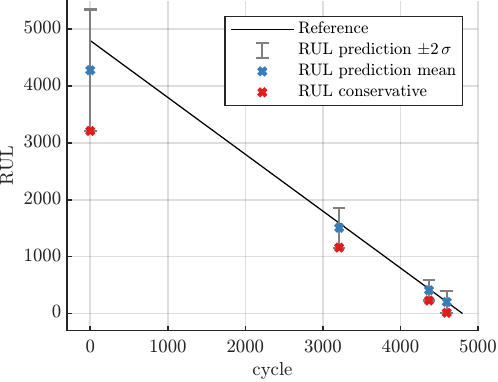}
     \caption{RUL monitoring strategy Cell 3}
    \label{fig:RUL_monitoringStrategy_example3}
\end{figure}
 \begin{figure}[h!]
     \centering
     \includegraphics[width=0.9\linewidth]{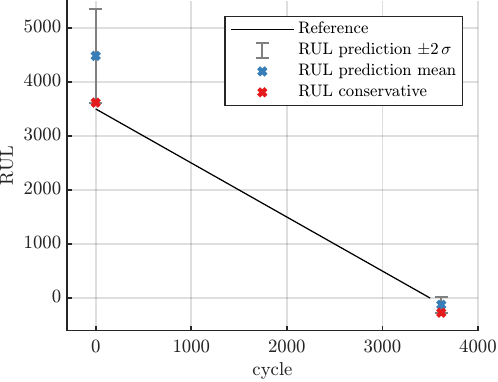}
     \caption{RUL monitoring strategy Cell 4}
    \label{fig:RUL_monitoringStrategy_example4}
\end{figure}
Observing the \gls{RUL} monitoring strategy from a cell behaving in an average manner in figure \ref{fig:RUL_monitoringStrategy_example3}, one notes, that the \gls{RUL} predictions nearly match the true value. Through the conservative selection of $k=2$, four monitoring steps are required. The operation cycles between two diagnostic charge cycles decreases monotonically with cell age. There is no overcycling, when \gls{EOL} of the monitoring strategy is reached. With $\Delta {N}_{\mathrm{EOL}} = 203$, already 95.8\% of the possible cycles were executed. If one looks at the fastest aging Cell 4 in figure \ref{fig:RUL_monitoringStrategy_example4}, the initial \gls{RUL} prediction noticeably deviate from the true values. Nevertheless, due to the conservative strategy, there is only a small overcycling of $\Delta {N}_{\mathrm{EOL}} = -116$ cycles.   
 \begin{figure}[h!]
     \centering
     \includegraphics[width=0.9\linewidth]{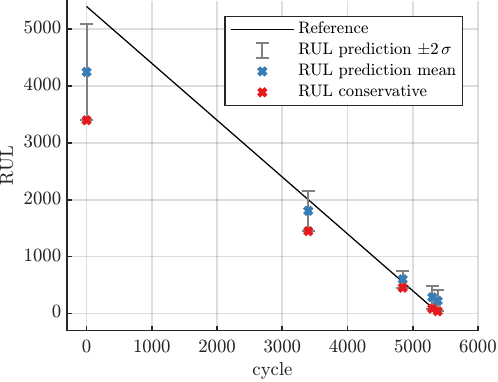}
     \caption{RUL monitoring strategy Cell 7}
    \label{fig:RUL_monitoringStrategy_example7}
\end{figure}
Taking a closer look at the slowest aging Cell 7 in figure \ref{fig:RUL_monitoringStrategy_example7}, consequently the initial \gls{RUL} prediction underestimates the battery life. From the second diagnostic cycle onwards, the \gls{RUL} is estimated properly. Here $\Delta {N}_{\mathrm{EOL}} = 17$ makes almost perfect use of the actual battery life, 99.7\% of the possible cycles were executed.
\begin{table}[ht]
    \centering
    \caption{RUL monitoring strategy results for $k=2$}
    \label{tab:RUL_monitoringStrategy}
    \renewcommand{\arraystretch}{1.15}
    \begin{tabular}{ccccccc}
        \hline
        \textbf{Cell} & $U$ & $M$ & $P_{\mathrm{over}}$ & $\Delta {N}_{\mathrm{EOL}}$ & $\Delta \mathrm{SoH}_{\mathrm{EOL}}$ \\
        \hline
        \textbf{1} & 0.9195 & 3 & 0 & 354  &  0.0104 \\
        \textbf{2} & 0.9717 & 4 & 0 & 119  &  0.0025 \\
        \textbf{3}  & 0.9577 & 4 & 0 & 203  &  0.0020 \\
        \textbf{4} & 1.0331 & 2 & 1 & -116 & -0.0074 \\
        \textbf{5} & 0.9370 & 3 & 0 & 271  &  0.0061 \\
        \textbf{6} & 0.9875 & 3 & 0 & 50   &  0.0008 \\
        \textbf{7} & 0.9969 & 5 & 0 & 17   &  0.0003 \\
        \textbf{8} & 0.9975 & 4 & 0 & 12   & -0.0052 \\
        \hline
    \end{tabular}
\end{table} 

For all eight cells the \glspl{KPI} of the \gls{RUL} monitoring strategy are summarized in table \ref{tab:RUL_monitoringStrategy}.
\section{Conclusion and Future Work}
\label{sec:Conclusion}
The presented work demonstrates that \gls{ICA}-based features, combined with data-efficient regression methods, enable accurate and uncertainty-aware estimation of both \gls{SoH} and \gls{RUL} for previously unseen lithium-ion cells using only single diagnostic charge cycles and aging data from a small set of similar reference cells. For \gls{SoH} estimation, the mapping from \gls{ICA} features to \gls{SoH} is found to be almost cell invariant, which leads to low \gls{NMAE} values around 1.3\% and allows even comparatively simple regression models such as \gls{SVR} and pooled \gls{GPR} to achieve highly satisfying accuracy with small maximum errors. In contrast, \gls{RUL} estimation exhibits pronounced cell-to-cell variance, resulting in a higher \gls{NMAE} of about 5.3\% and clearly revealing the need for models that explicitly account for heterogeneous degradation trajectories. Here, the \gls{GPRn} ensemble successfully averages over individual cell models, reduces overfitting compared to pooled \gls{GPR}, and provides a meaningful decomposition into epistemic and aleatoric uncertainty, with the latter dominating due to intrinsic cell-to-cell variability.

Building on these uncertainty-aware \gls{RUL} estimates, the proposed monitoring strategy shows that a battery can be safely operated from \gls{BOL} to \gls{EOL} by performing only a few (typically three to five) diagnostic cycles, while exploiting more than 90\% of the available cycle budget and keeping \gls{SoH} deviations at strategy \gls{EOL} well below 1\% for most cells. By tuning the strategy parameter $k$, operators can trade off utilization grade, number of monitoring steps, and overcycling risk, which makes the approach directly applicable to scenarios with sparse measurements, such as second-life deployment after unknown first-life usage.

Future work can extend the current \gls{RUL} monitoring strategy, where predictions only rely on a single-cycle measurement. By incorporating information from multiple already taken diagnostic cycles of the same test cell instead of ignoring them, for example by storing past \gls{SoH} estimates, degradation trends can be constructed. If there is a bigger dataset available, the degradation trend of the currently monitored cell can be compared to degradation trends of all training cells. Using this information  one can adapt the training cell weights $\pi_i$ in Eq. (\ref{eq:pi}) within the \gls{GPRn} ensemble. This concept could reduce the dominance of aleatoric uncertainty in the \gls{RUL} estimation. Success depends heavily on the size of the data set. Eight cells are far from being sufficient to describe cell variant aging trends in a statistically reliable manner.  

Generally, also the progression of the \gls{ICA} features on aging cells  under varying operating conditions should be investigated. Incorporating different charging rates, discharging profiles and ambient temperatures generalizes the presented method for more practical applications.

\balance
\interlinepenalty=10000
\bibliographystyle{apacite}

@Inbook{Spearman,
title={Spearman Rank Correlation Coefficient},
bookTitle={The Concise Encyclopedia of Statistics},
year={2008},
publisher={Springer New York},
address={New York, NY},
pages={502--505},
XXXisbn={978-0-387-32833-1},
XXXdoi={10.1007/978-0-387-32833-1_379},
XXXurl={https://doi.org/10.1007/978-0-387-32833-1_379},
}

@article{guo_review_2021,
	title = {A {Review} of {Equivalent} {Circuit} {Model} {Based} {Online} {State} of {Power} {Estimation} for {Lithium}-{Ion} {Batteries} in {Electric} {Vehicles}},
	volume = {4},
	issn = {2624-8921},
	XXXurl = {https://www.mdpi.com/2624-8921/4/1/1},
	XXXdoi = {10.3390/vehicles4010001},
	language = {en},
	number = {1},
	XXXurldate = {2025-12-15},
	journal = {Vehicles},
	author = {Guo, Ruohan and Shen, Weixiang},
	month = dec,
	year = {2021},
	pages = {1--29},
}

@article{jossen_fundamentals_2006,
	title = {Fundamentals of battery dynamics},
	volume = {154},
	copyright = {https://www.elsevier.com/tdm/userlicense/1.0/},
	issn = {03787753},
	XXXurl = {https://linkinghub.elsevier.com/retrieve/pii/S0378775305014321},
	XXXdoi = {10.1016/j.jpowsour.2005.10.041},
	language = {en},
	number = {2},
	XXXurldate = {2025-12-15},
	journal = {Journal of Power Sources},
	author = {Jossen, Andreas},
	month = mar,
	year = {2006},
	pages = {530--538},
}

@phdthesis{schmitt_state_2022,
	address = {München},
	title = {State of health estimation for lithium-ion batteries},
	language = {de},
	school = {TUM School of Engineering and Design der Technischen Universität München},
	author = {Schmitt, Julius},
	month = dec,
	year = {2022},
    howpublished = {PhD thesis, Technical University of Munich},
}

@phdthesis{hamar_data-driven_2024,
	address = {München},
	title = {Data-driven battery state estimation},
	language = {en},
	school = {Technische Universität München  TUM School of Engineering and Design},
	author = {Hamar, Jacob Clay},
	month = may,
	year = {2024},
    howpublished = {PhD thesis, Technical University of Munich},
}

@article{al-hashimi_decade_2025,
	title = {A decade of machine learning in lithium-ion battery state estimation: a systematic review},
	volume = {31},
	issn = {0947-7047, 1862-0760},
	shorttitle = {A decade of machine learning in lithium-ion battery state estimation},
	XXXurl = {https://link.springer.com/10.1007/s11581-024-06049-4},
	XXXdoi = {10.1007/s11581-024-06049-4},
	language = {en},
	number = {3},
	XXXurldate = {2025-12-15},
	journal = {Ionics},
	author = {Al-Hashimi, Zaina and Khamis, Taha and Al Kouzbary, Mouaz and Arifin, Nooranida and Mokayed, Hamam and Abu Osman, Noor Azuan},
	month = mar,
	year = {2025},
	pages = {2351--2377},
}

@article{dar_comprehensive_2025,
	title = {A comprehensive review, perspectives and future directions of battery characterization and parameter estimation},
	volume = {55},
	issn = {0021-891X, 1572-8838},
	XXXurl = {https://link.springer.com/10.1007/s10800-024-02217-6},
	XXXdoi = {10.1007/s10800-024-02217-6},
	language = {en},
	number = {4},
	XXXurldate = {2025-12-15},
	journal = {Journal of Applied Electrochemistry},
	author = {Dar, Tasadeek Hassan and Singh, Satyavir},
	month = apr,
	year = {2025},
	pages = {837--863},
}

@article{demirci_review_2024,
	title = {Review of battery state estimation methods for electric vehicles-{Part} {II}: {SOH} estimation},
	volume = {96},
	issn = {2352152X},
	shorttitle = {Review of battery state estimation methods for electric vehicles-{Part} {II}},
	XXXurl = {https://linkinghub.elsevier.com/retrieve/pii/S2352152X24022898},
	XXXdoi = {10.1016/j.est.2024.112703},
	language = {en},
	XXXurldate = {2025-12-15},
	journal = {Journal of Energy Storage},
	author = {Demirci, Osman and Taskin, Sezai and Schaltz, Erik and Acar Demirci, Burcu},
	month = aug,
	year = {2024},
	pages = {112703},
}

@article{yao_review_2021,
	title = {A {Review} of {Lithium}-{Ion} {Battery} {State} of {Health} {Estimation} and {Prediction} {Methods}},
	volume = {12},
	issn = {2032-6653},
	XXXurl = {https://www.mdpi.com/2032-6653/12/3/113},
	XXXdoi = {10.3390/wevj12030113},
	language = {en},
	number = {3},
	XXXurldate = {2025-12-15},
	journal = {World Electric Vehicle Journal},
	author = {Yao, Lei and Xu, Shiming and Tang, Aihua and Zhou, Fang and Hou, Junjian and Xiao, Yanqiu and Fu, Zhijun},
	month = aug,
	year = {2021},
	pages = {113},
}

@article{wang_state--health_2023,
	title = {A state-of-health estimation method based on incremental capacity analysis for {Li}-ion battery considering charging/discharging rate},
	volume = {73},
	issn = {2352152X},
	XXXurl = {https://linkinghub.elsevier.com/retrieve/pii/S2352152X23024088},
	XXXdoi = {10.1016/j.est.2023.109010},
	language = {en},
	XXXurldate = {2025-12-15},
	journal = {Journal of Energy Storage},
	author = {Wang, Guangfeng and Cui, Naxin and Li, Changlong and Cui, Zhongrui and Yuan, Haitao},
	month = dec,
	year = {2023},
	pages = {109010},
}

@article{beatty_review_2024,
	title = {A {Review} of {Methods} of {Generating} {Incremental} {Capacity}–{Differential} {Voltage} {Curves} for {Battery} {Health} {Determination}},
	volume = {17},
	issn = {1996-1073},
	XXXurl = {https://www.mdpi.com/1996-1073/17/17/4309},
	XXXdoi = {10.3390/en17174309},
	language = {en},
	number = {17},
	XXXurldate = {2025-12-15},
	journal = {Energies},
	author = {Beatty, Matthew and Strickland, Dani and Ferreira, Pedro},
	month = aug,
	year = {2024},
	pages = {4309},
}

@article{chen_filter_2024,
	title = {Filter methods comparation for incremental capacity analysis in lithium-ion batteries health prediction},
	volume = {101},
	issn = {2352152X},
	XXXurl = {https://linkinghub.elsevier.com/retrieve/pii/S2352152X24034649},
	XXXdoi = {10.1016/j.est.2024.113878},
	language = {en},
	XXXurldate = {2025-12-15},
	journal = {Journal of Energy Storage},
	author = {Chen, Hongxing and Yue, Wenhui and Bin, Guangfu and Jiang, Qi and Shao, Wei and She, Chengqi},
	month = nov,
	year = {2024},
	pages = {113878},
}

@article{stroe_lithium-ion_2020,
	title = {Lithium-{Ion} {Battery} {State}-of-{Health} {Estimation} {Using} the {Incremental} {Capacity} {Analysis} {Technique}},
	volume = {56},
	copyright = {https://ieeexplore.ieee.org/Xplorehelp/downloads/license-information/IEEE.html},
	issn = {0093-9994, 1939-9367},
	XXXurl = {https://ieeexplore.ieee.org/document/8911243/},
	XXXdoi = {10.1109/TIA.2019.2955396},
	language = {en},
	number = {1},
	XXXurldate = {2025-12-15},
	journal = {IEEE Transactions on Industry Applications},
	author = {Stroe, Daniel-Ioan and Schaltz, Erik},
	month = jan,
	year = {2020},
	pages = {678--685},
}

@article{li_state_2020,
	title = {State of health estimation for {Li}-{Ion} battery using incremental capacity analysis and {Gaussian} process regression},
	volume = {190},
	issn = {03605442},
	XXXurl = {https://linkinghub.elsevier.com/retrieve/pii/S0360544219321620},
	XXXdoi = {10.1016/j.energy.2019.116467},
	language = {en},
	XXXurldate = {2025-12-15},
	journal = {Energy},
	author = {Li, Xiaoyu and Yuan, Changgui and Li, Xiaohui and Wang, Zhenpo},
	month = jan,
	year = {2020},
	pages = {116467},
}

@book{rasmussen_gaussian_2008,
	address = {Cambridge, Mass.},
	edition = {3. print},
	series = {Adaptive computation and machine learning},
	title = {Gaussian processes for machine learning},
	isbn = {978-0-262-18253-9},
	language = {en},
	publisher = {MIT Press},
	author = {Rasmussen, Carl Edward and Williams, Christopher K. I.},
	year = {2008},
}

@article{li_state_2020-1,
	title = {State of health estimation for {Li}-ion battery via partial incremental capacity analysis based on support vector regression},
	volume = {203},
	issn = {03605442},
	XXXurl = {https://linkinghub.elsevier.com/retrieve/pii/S0360544220309592},
	XXXdoi = {10.1016/j.energy.2020.117852},
	language = {en},
	XXXurldate = {2025-12-15},
	journal = {Energy},
	author = {Li, Xiaoyu and Yuan, Changgui and Wang, Zhenpo},
	month = jul,
	year = {2020},
	pages = {117852},
}

@article{pang_lithium-ion_2021,
	title = {A lithium-ion battery remaining useful life prediction method based on the incremental capacity analysis and {Gaussian} process regression},
	volume = {127},
	issn = {00262714},
	XXXurl = {https://linkinghub.elsevier.com/retrieve/pii/S0026271421003711},
	XXXdoi = {10.1016/j.microrel.2021.114405},
	language = {en},
	XXXurldate = {2025-12-15},
	journal = {Microelectronics Reliability},
	author = {Pang, Xiaoqiong and Liu, Xiaoyan and Jia, Jianfang and Wen, Jie and Shi, Yuanhao and Zeng, Jianchao and Zhao, Zhen},
	month = dec,
	year = {2021},
	pages = {114405},
}

@article{xia_state_2023,
	title = {State of health and remaining useful life prediction of lithium-ion batteries based on a disturbance-free incremental capacity and differential voltage analysis method},
	volume = {64},
	issn = {2352152X},
	XXXurl = {https://linkinghub.elsevier.com/retrieve/pii/S2352152X23005583},
	XXXdoi = {10.1016/j.est.2023.107161},
	language = {en},
	XXXurldate = {2025-12-15},
	journal = {Journal of Energy Storage},
	author = {Xia, Fei and Wang, Kangan and Chen, Jiajun},
	month = aug,
	year = {2023},
	pages = {107161},
}

@inproceedings{jha_itransformer_2024,
	address = {Chicago, IL, USA},
	title = {{iTransformer} {Network} {Based} {Approach} for {Accurate} {Remaining} {Useful} {Life} {Prediction} in {Lithium}-{Ion} {Batteries}},
	copyright = {https://XXXdoi.org/10.15223/policy-029},
	isbn = {979-8-3503-1766-4},
	XXXurl = {https://ieeexplore.ieee.org/document/10598898/},
	XXXdoi = {10.1109/ITEC60657.2024.10598898},
	language = {en},
	XXXurldate = {2025-12-15},
	booktitle = {2024 {IEEE} {Transportation} {Electrification} {Conference} and {Expo} ({ITEC})},
	publisher = {IEEE},
	author = {Jha, Anurag and Dorkar, Oorja and Biswas, Atriya and Emadi, Ali},
	month = jun,
	year = {2024},
	pages = {1--8},
}

@article{elmahallawy_comprehensive_2022,
	title = {A {Comprehensive} {Review} of {Lithium}-{Ion} {Batteries} {Modeling}, and {State} of {Health} and {Remaining} {Useful} {Lifetime} {Prediction}},
	volume = {10},
	copyright = {https://creativecommons.org/licenses/by/4.0/legalcode},
	issn = {2169-3536},
	XXXurl = {https://ieeexplore.ieee.org/document/9944663/},
	XXXdoi = {10.1109/ACCESS.2022.3221137},
	language = {en},
	XXXurldate = {2025-12-18},
	journal = {IEEE Access},
	author = {Elmahallawy, Mohamed and Elfouly, Tarek and Alouani, Ali and Massoud, Ahmed M.},
	year = {2022},
	pages = {119040--119070},
}

@article{santhanagopalan_review_2006,
	title = {Review of models for predicting the cycling performance of lithium ion batteries},
	volume = {156},
	copyright = {https://www.elsevier.com/tdm/userlicense/1.0/},
	issn = {03787753},
	XXXurl = {https://linkinghub.elsevier.com/retrieve/pii/S0378775305007810},
	XXXdoi = {10.1016/j.jpowsour.2005.05.070},
	language = {en},
	number = {2},
	XXXurldate = {2025-12-18},
	journal = {Journal of Power Sources},
	author = {Santhanagopalan, Shriram and Guo, Qingzhi and Ramadass, Premanand and White, Ralph E.},
	month = jun,
	year = {2006},
	pages = {620--628},
}

@misc{zhou_battery_2025,
	title = {Battery {State} of {Health} {Estimation} and {Incremental} {Capacity} {Analysis} under {General} {Charging} {Profiles} {Using} {Neural} {Networks}},
	XXXurl = {http://arxiv.org/abs/2502.19586},
	XXXdoi = {10.48550/arXiv.2502.19586},
	language = {en},
	XXXurldate = {2025-12-18},
	author = {Zhou, Qinan and Vuylsteke, Gabrielle and Anderson, R. Dyche and Sun, Jing},
	month = jun,
	year = {2025},
	note = {arXiv:2502.19586 [eess]},
}

@article{she_battery_2023,
	title = {Battery {State}-of-{Health} {Estimation} {Based} on {Incremental} {Capacity} {Analysis} {Method}: {Synthesizing} {From} {Cell}-{Level} {Test} to {Real}-{World} {Application}},
	volume = {11},
	copyright = {https://ieeexplore.ieee.org/Xplorehelp/downloads/license-information/IEEE.html},
	issn = {2168-6777, 2168-6785},
	shorttitle = {Battery {State}-of-{Health} {Estimation} {Based} on {Incremental} {Capacity} {Analysis} {Method}},
	XXXurl = {https://ieeexplore.ieee.org/document/9537755/},
	XXXdoi = {10.1109/JESTPE.2021.3112754},
	language = {en},
	number = {1},
	XXXurldate = {2025-12-18},
	journal = {IEEE Journal of Emerging and Selected Topics in Power Electronics},
	author = {She, Chengqi and Zhang, Lei and Wang, Zhenpo and Sun, Fengchun and Liu, Peng and Song, Chunbao},
	month = feb,
	year = {2023},
	pages = {214--223},
}

@article{ansean_lithium-ion_2019,
	title = {Lithium-{Ion} {Battery} {Degradation} {Indicators} {Via} {Incremental} {Capacity} {Analysis}},
	volume = {55},
	copyright = {https://ieeexplore.ieee.org/Xplorehelp/downloads/license-information/IEEE.html},
	issn = {0093-9994, 1939-9367},
	XXXurl = {https://ieeexplore.ieee.org/document/8603757/},
	XXXdoi = {10.1109/TIA.2019.2891213},
	language = {en},
	number = {3},
	XXXurldate = {2025-12-18},
	journal = {IEEE Transactions on Industry Applications},
	author = {Ansean, David and Garcia, Victor Manuel and Gonzalez, Manuela and Blanco-Viejo, Cecilio and Viera, Juan Carlos and Pulido, Yoana Fernandez and Sanchez, Luciano},
	month = may,
	year = {2019},
	pages = {2992--3002},
}

@article{nguyen_van_estimation_2023,
	title = {Estimation of {SoH} and internal resistances of {Lithium} ion battery based on {LSTM} network},
	volume = {18},
	issn = {14523981},
	XXXurl = {https://linkinghub.elsevier.com/retrieve/pii/S1452398123001931},
	XXXdoi = {10.1016/j.ijoes.2023.100166},
	language = {en},
	number = {6},
	XXXurldate = {2025-12-18},
	journal = {International Journal of Electrochemical Science},
	author = {Nguyen Van, Chi and Quang, Duy Ta},
	month = jun,
	year = {2023},
	pages = {100166},
}

@phdthesis{birkl_diagnosis_2016,
	address = {Oxford},
	title = {Diagnosis and {Prognosis} of {Degradation} in {Lithium}-{Ion} {Batteries}},
	language = {en},
	school = {University of Oxford},
	author = {Birkl, Christoph},
	year = {2016},
}

@misc{birkl_oxford_2017,
	title = {Oxford {Battery} {Degradation} {Dataset} 1},
	copyright = {ODC Attribution Share-Alike for data/databases (ODC-ODbL)},
	XXXurl = {http://ora.ox.ac.uk/objects/uuid:03ba4b01-cfed-46d3-9b1a-7d4a7bdf6fac},
	XXXdoi = {10.5287/BODLEIAN:KO2KDMYGG},
	language = {en},
	XXXurldate = {2026-01-07},
	publisher = {University of Oxford},
	author = {Birkl, Christoph},
	collaborator = {Howey, David},
	year = {2017},
}

@article{goldberg_eigentaste_2001,
	title = {Eigentaste: {A} {Constant} {Time} {Collaborative} {Filtering} {Algorithm}},
	volume = {4},
	issn = {1573-7659},
	XXXurl = {https://doi.org/10.1023/A:1011419012209},
	XXXdoi = {10.1023/A:1011419012209},
	number = {2},
	journal = {Information Retrieval},
	author = {Goldberg, Ken and Roeder, Theresa and Gupta, Dhruv and Perkins, Chris},
	month = jul,
	year = {2001},
	pages = {133--151},
}
\PHMbibliography{literature}

\end{document}